# A PEDAGOGICAL DISCUSSION CONCERNING THE GRAVITATIONAL ENERGY RADIATED BY KEPLERIAN SYSTEMS

Christian Bracco
UMR Fizeau, Université de Nice-Sophia Antipolis, CNRS, Observatoire de la Côte d'Azur, Campus Valrose, F-06108 Nice Cedex and
Syrte, CNRS, Observatoire de Paris, 61 avenue de l'Observatoire, F-75014 Paris

Jean-Pierre Provost
INLN, Université de Nice-Sophia Antipolis, 1361 route des lucioles, Sophia Antipolis, F-06560 Valbonne

Pierre Salati
Laboratoire d'Annecy-Le-Vieux de Physique Théorique LAPTH, Université de Savoie et CNRS, 9 Chemin de Bellevue, B.P. 110, F-74941 Annecy-Le-Vieux Cedex

We first discuss the use of dimensional arguments (and of the quadrupolar emission hypothesis) in the derivation of the gravitational power radiated on a circular orbit. Then, we show how to simply obtain the instantaneous power radiated on a general Keplerian orbit by approximating it locally by a circle. This allows recovering with a good precision, in the case of an ellipse, the highly non trivial dependence on the eccentricity of the average power given by general relativity. The whole approach is understandable by undergraduate students.

**Keywords:** gravitational waves, Keplerian orbits, PSR 1913+16, dimensional analysis, radius of curvature

**PACS:** 04.30.-w, 95.85.Sz

## I. GRAVITATIONAL ENERGY RADIATED BY KEPLERIAN SYSTEMS; INTRODUCTION

Einstein built the general theory of relativity (GR) between 1907, when he formulated the first version of the equivalence principle, and November 1915, when he obtained his equations for the gravitational field. In GR, space-time is described by the metric tensor, whose components are identified with the gravitational potentials of the matter. In 1918, Einstein established that small perturbations of this tensor propagate at the speed of light and are generated by masses undergoing acceleration. These perturbations, which are called gravitational waves (GW) describe transverse (shear) deformations of space and are associated with the quadrupolar momentum (inertia momentum) of the source[1,2] (whereas in electromagnetism the transverse polarization is a vectorial one and is generated in first approximation by the dipolar momentum of the source).

In the case of a binary system, the energy carried away by GW is lost by the system, hence a decrease of its orbital period. The first detection of that effect occurred in the 1980's after the discovery in 1975[3] of the binary pulsar PSR 1913+16 by Hulse and Taylor, who were rewarded by the Nobel Prize in 1993.[4] This system is composed of two neutron stars with 1.44 and 1.38 solar masses, orbiting with a short period of 7.8 h on Keplerian elliptical trajectories with a noticeable eccentricity 0.617. Thanks to many years of observations by Taylor and Weisberg, the data were accurate enough to show that the decrease of the orbital period is consistent, up to a precision of measurement of 0.4%, with the emission of GW predicted by GR.[5] GW detection is an intense field of research, and huge detectors, such as the LIGO and VIRGO (or LISA in the future) interferometers[6] have been built to detect GW as they reach the Earth.

The aim of this paper is to show that standard knowledge at undergraduate level can be used to calculate the GW power radiated by a celestial body such as PSR 1913+16, and to discuss its dependence on the various orbital parameters (including eccentricity). In section II, we use dimensional arguments to obtain (up to a constant) the gravitational power radiated on a circular orbit under the assumption of a quadrupolar emission. The same analysis, applied to the dipolar and quadrupolar electromagnetic radiations leads to an interesting comparison. Of course, this dimensional approach does not apply to a general Keplerian orbit because of the eccentricity, which is an adimensional parameter. In section III, we show how to derive very simply the instantaneous power of radiation on such an orbit, by introducing the local radius of curvature (which enters the well known expression of the normal acceleration). Finally in section IV, we discuss the case of a Keplerian ellipse and calculate the mean radiated power. The dependence of this power with respect to the eccentricity of the orbit agrees with a precision of order 1% with the formula[7] of GR which has been confronted to the observations.

**II. DIMENSIONAL CALCULUS AND THE CASE OF A CIRCULAR ORBIT**

In the center of mass frame of a binary system, let $\mu = m_1 m_2/(m_1 + m_2)$ be the reduced mass, $R$ be the radius of its circular orbit and $T = 2\pi/\omega$ be its orbital period. The power $P_{\text{circ}}$ of the gravitational radiation is expected to depend on $\mu$, $R$ and $\omega$, as well as on the coupling constant of gravitation $G$ and on the speed of light $c$. It is impossible to obtain a unique formula for $P_{\text{circ}}$ from a dimensional argument with these five parameters, since there are only three dimensions, say the mass $M$, the length $L$ and the speed $V$ (or the time $T$). The problem can actually be solved if we remember that GW are shear perturbations and consequently have a quadrupolar origin. The wave amplitude is then proportional to $\mu R^2$ (inertia momentum) and $P_{\text{circ}}$ to its square $\mu^2 R^4$. Writing the dimensional equality

$$[P_{\text{circ}}] \equiv M^2 L^4 [G]^\alpha [\omega]^\beta [V]^\gamma = ML^{-1}V^3, \tag{1}$$

and using the relations $[G] \equiv M^{-1}LV^2$, $[c] \equiv V$, $[\omega] \equiv L^{-1}V$, one gets by identification $\alpha = 1, \beta = 6, \gamma = -5$, and:

$$P_{\text{circ}} = AG\mu^2 c^{-5} \omega^6 R^4. \tag{2}$$

In Eq. (2), $A$ is a constant. Its exact value $32/5$ is not far from unity[8] and is given by a rather tedious calculation in GR (see Ref. 1-2).

Let us emphasize that the $\omega^6$ dependence of $P_{\text{circ}}$ is characteristic of a quadrupolar radiation. Indeed, as soon as we expect $P_{\text{circ}}$ to be proportional to $\mu^2$, Eq. (1) implies that it must also be proportional to $G$, and that the powers of $L$ and $\omega$ are directly related; more precisely $P_{\text{circ}}$ becomes equal to $AG\mu^2 \omega^{2(n+1)} R^{2n} c^{-(2n+1)}$ where $n$ is an integer describing the multipolar order of the source. Such an analysis also applies in electromagnetism. There $P_{\text{circ}}$ is obtained from its proportionality to the square $q^2$ of the charge and to appropriate powers of $R$, $\omega$ and of the constants $c$ and $4\pi\varepsilon_0$. The dimensional equality $[P_{\text{circ}}] \equiv Q^2 L^{2n} [4\pi\varepsilon_0]^\alpha [\omega]^\beta [V]^\gamma = ML^{-1}V^3$ and the relation $[4\pi\varepsilon_0]^{-1} = MLV^2 Q^{-2}$ lead to the expression $P_{\text{circ}} = Aq^2 (4\pi\varepsilon_0)^{-1} R^{2n} \omega^{2(n+1)} c^{-2n+1}$. In the familiar dipolar case[9] ($n=1$), the radiated power is proportional to $\omega^4$, the exact calculation giving $A = 2/3$.

**III. RADIUS OF CURVATURE AND THE EXTENSION TO KEPLERIAN ORBITS**

The above dimensional argument[10] cannot be applied to an elliptical trajectory because eccentricity is an adimensional parameter. A way to bypass this difficulty is to consider that locally (near a point $M$) the trajectory can be approximated (up to second order) by a circle whose radius $R = MC$ is called the local radius of curvature (figure 1). Then, the instantaneous angular velocity $\omega$ is $\omega = V/R$, where $V$ is the velocity at the point $M$. This allows writing the instantaneous power of gravitational radiation as:

$$P = \frac{dW}{dt} = AG\mu^2 c^{-5} V^6 R^{-2}. \tag{3}$$

In kinematics, the radius of curvature is also well known from the relation $a_N = V^2/R \, (= \omega^2 R)$, which gives the normal acceleration, i.e. the projection of the acceleration $\boldsymbol{a}$ on the line $MC$. Let $\varphi$ be the angle between $\boldsymbol{a}$ and $MC$, then:

$$\frac{V^2}{R} = \|\boldsymbol{a}\| \cos \varphi. \tag{4}$$

When the trajectory is governed by central forces pointing towards a point $O$ (figure 1), $\varphi$ is also the angle between the velocity and the normal to the line $OM$. In that case, the conservation of the angular momentum $\mu \boldsymbol{r} \wedge \boldsymbol{V}$ leads to the additional relation

$$rV\cos\varphi = K \qquad (r = OM, \ K \text{ area's constant}), \tag{5}$$

and the product of Eq. (4) and Eq. (5) leads to the factor[11] $V^6 R^{-2}$ of Eq. 3, through the term

$$\frac{V^3}{R} = \frac{K\|\boldsymbol{a}\|}{r}. \tag{6}$$

Finally, in the case of a Keplerian motion, where

$$\|\boldsymbol{a}\| = \alpha r^{-2} \qquad (\alpha = G(m_1 + m_2)), \tag{7}$$

the instantaneous power of radiation becomes:

$$P = \frac{dW}{dt} = \frac{AG\mu^2 c^{-5} \alpha^2 K^2}{r^6} \ . \tag{8}$$

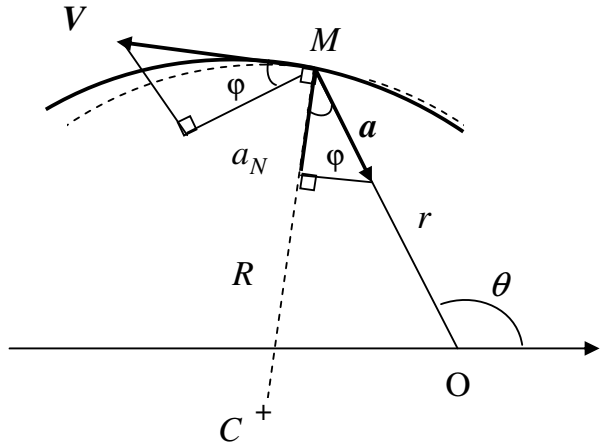

**Fig. 1. Polar coordinates and kinematics in the case of a centripetal acceleration.** The arc of trajectory (bold curve) is approximated at *M* by a circle of center *C* and radius *R* (dotted curve). $\varphi$ is the angle between *MC* (perpendicular to the velocity *V*) and *OM* (parallel to the acceleration *a*). It is also the angle between *V* and the normal to *OM*.

In the discussion of the next section, we shall also need the expression

$$\frac{dW}{d\theta} = \frac{AG\mu^2 c^{-5} \alpha^2 K}{r^4}, \tag{9}$$

which gives the energy d*W* radiated along an arc d$\theta$ of the trajectory. It is a consequence of Eq. (8) and of the relation (area's law):

$$K = r^2 \frac{d\theta}{dt} \ . \tag{10}$$

## IV. THE INFLUENCE OF THE ECCENTRICITY FOR AN ELLIPTICAL ORBIT

We now examine some consequences of Eqs (8) and (9) in the case of a Keplerian ellipse. Let us recall that in polar coordinates, with origin $O$ at a focus (figure 2) and with $\theta$ the true anomaly, its equation reads: [12]

$$r = \frac{p}{(1+e\cos\theta)} \quad \text{with} \quad p = K^2/\alpha, \quad 0 \le e < 1. \tag{11}$$

In the following, we shall need the relation

$$p = a(1-e^2), \tag{12}$$

between the parameter $p$, the eccentricity $e$ and the semi-major axis $a$ of the ellipse, as well as the expression of the period of the motion $T$ (Kepler's third law):

$$T = 2\pi a^{3/2} \alpha^{-1/2} \quad \text{or} \quad \omega^2 a^3 = \alpha. \tag{13}$$

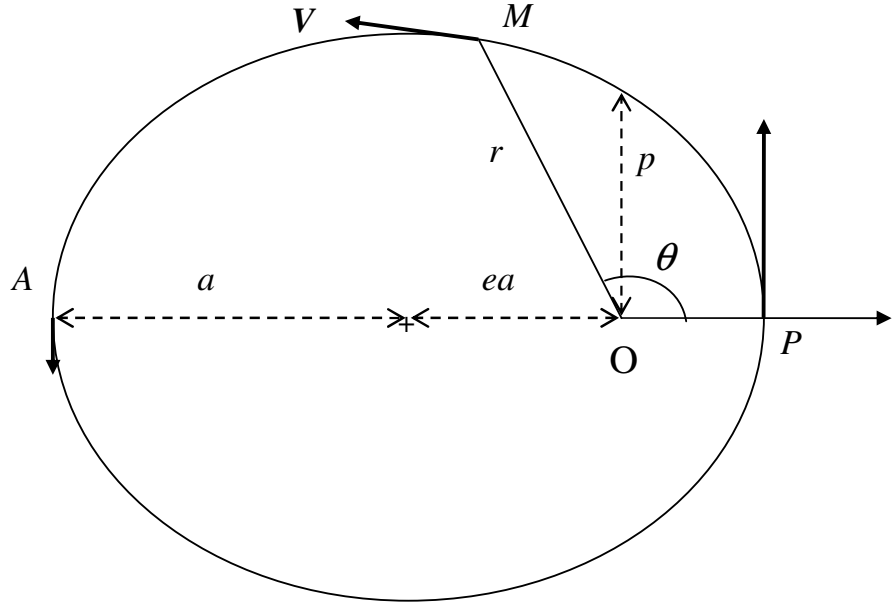

**Fig. 2. A Keplerian ellipse with $e = 0.6$.** The origin $O$ is a focus and $\theta$ is the true anomaly. $P$ is the periastron $(OP = a(1-e))$, $A$ the apastron $(OA = a(1+e))$ and $AP = 2a$ the major axis. The velocity $V$ obeys the area's law.

First, we remark from Eq. (8) (and Eq. (11)) that when the eccentricity is not negligible, the ratio $((1+e)/(1-e))^6$ of the instantaneous powers radiated at the periastron $P$ $(\theta=0)$ and at the apastron $A$ $(\theta=\pi)$ may be very large, e.g. 4000 for $e=0.6$. This suggests that the radiation is then mainly emitted near the periastron. One can check it by using Eq. (9) to obtain by a direct numerical integration the angular arc which corresponds to 50% of the radiated energy. For $e=0.6$ (figure 2), this arc is about 1 rd (centered on $P$) and it corresponds to a small part of the orbit. The associated travel time is even smaller compared to the period, approximately 4%, because of the area's law. (For $e=0.9$, these values are respectively 0.9 rd and 0.6% $T$). On figure 3 we have represented in the interval $[-\pi;\pi]$, both $dW/d\theta$ (solid curve) and $P=dW/dt$ (dotted long dashed line) respectively as a function of the true anomaly $\theta$ and of the mean anomaly $2\pi t/T$. These functions are normalized to their maximum value at the periastron $P$ ($\theta=0, t=0$).

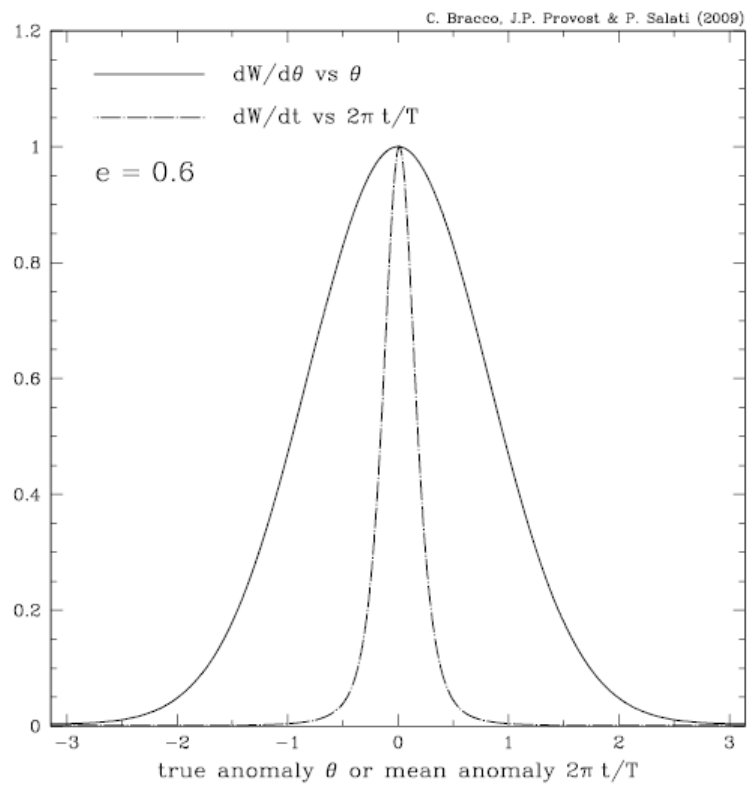

**Fig. 3. Angular and time repartition of the GW radiated energy.**

Another remark concerns the dependence with respect to the eccentricity of the time averaged power $\bar{P}$ (total energy radiated on the orbit divided by the period $T = 2\pi/\omega$):

$$\bar{P} = \frac{\omega}{2\pi} \int_0^{2\pi} \frac{dW}{d\theta} d\theta \ . \tag{14}$$

Using Eq. (9) for $dW/d\theta$ and the relations

$$\frac{\alpha^2 K}{p^4} = \alpha^{5/2} p^{-7/2} = \omega^5 a^4 \left(1-e^2\right)^{-7/2} \ , \tag{15}$$

which follow respectively from Eq. (11) and Eqs. (12)-(13), one immediately gets:

$$\bar{P} = AG\mu^2 c^{-5}\omega^6 a^4 \times \left(1-e^2\right)^{-7/2} \times \frac{1}{2\pi}\int_0^{2\pi} (1+e\cos\theta)^4 d\theta. \tag{16}$$

For our discussion, we write Eq. (16) as

$$\bar{P} = P_{circ} \times f(e) \ , \tag{17}$$

with (after the evaluation of the angular integral):

$$P_{circ} = AG\mu^2 c^{-5}\omega^6 a^4 \ ; \quad f(e) = \left(1-e^2\right)^{-7/2}\left(1+3e^2+\frac{3}{8}e^4\right) \ . \tag{18}$$

$P_{circ}$ of Eq. (18) is the power which would be radiated on a circular orbit of radius $a$. The function $f(e)$ exhibits the influence of the eccentricity. Of course, for $e = 0$ (circular orbit), we recover expression (2) of $P_{circ}$ as expected. In the general case $e \neq 0$, $f(e)$ is a magnifying factor[13] with respect to the power radiated on a circular orbit of radius $a$. It is approximately 11.8 in the case of PSR 1913+16. More generally, the factor $\left(1-e^2\right)^{-7/2}$ in $f(e)$ implies that the power $\bar{P}$ of the GW emission increases significantly when $e \to 1$, i.e. for high eccentric orbits ($a$ being fixed).

As a final remark, let us recall that the above calculation of $\bar{P}$ is based on an approximation (assimilating locally the trajectory to a circle). A more exact formula (see Ref. 7), which has been confronted to the observations of PSR 1913+16, differs from Eq. (18) by the expression of the magnifying factor:

$$f^*(e) = \left(1-e^2\right)^{-7/2} \times \left(1 + \frac{73}{24}e^2 + \frac{37}{96}e^4\right). \tag{19}$$

However, since the factors of $e^2$ and $e^4$ in the last parenthesis of Eq. (18) are $3 = 72/24$ and $3/8 = 36/96$, the expressions of $f(e)$ and $f^*(e)$ are quite close. In particular, they share the same $\left(1-e^2\right)^{-7/2}$ critical dependence on the eccentricity. In the case of PSR 1913+16, the relative discrepancy between them is of the order of 0.8%, i.e. only twice the precision of the measurements. In the general case, the error does not exceed 1.2%.

In conclusion, the formula obtained in section II for the power radiated by GW on a circular orbit, and the introduction of the local radius of curvature, have led us to recover the non trivial influence of the eccentricity given by GR in the realistic case of an elliptical orbit. The whole calculation can be considered as an application at the undergraduate level (dimensional analysis, normal acceleration and Kepler's laws) of a problem of high physical interest.

---

[1] Steven Weinberg, *Gravitation and Cosmology* (John Wiley and Sons, New York, 1972)

[2] Charles W Misner, Kip S Thorne, and John Archibald Wheeler, *Gravitation* (W. H. Freeman, San Francisco, 1973)

[3] Russell A Hulse and Joseph H Taylor, Astrophys. J **195**, L51-L53 (1975).

[4] see Joseph H Taylor and Russell A Hulse Nobel lectures <http://nobelprize.org/nobel_prizes/physics/laureates/1993/index.html>

[5] Joseph H Taylor and Joel M Weisberg, Astrophys. J. **253**, 908-920 (1982); Joseph H Taylor and Joel M Weisberg, Astrophys. J. **345**, 434-450 (1989); Thibault Damour and Joseph H Taylor, Phys. Rev. D **45**, 1840-1868 (1992).

[6] see e.g. LIGO <http://www.ligo.caltech.edu> , VIRGO < http://www.virgo.infn.it > and LISA < http://lisa.nasa.gov/> homepages.

[7] C Peters and J Mathews, Phys. Rev. **131**, 435-440 (1963).

[8] Had we taken $[T]^{-\beta}$ instead of $[\omega]^{\beta}$ in the dimensional equality, the constant would have been multiplied by a very large factor $(2\pi)^6 \sim 6.10^4$ and the dimensional analysis would have been of little interest to derive orders of magnitude. The choice of $\omega^{-1}$ instead of $T$ for the characteristic time of evolution of harmonic phenomena is discussed for example in J.-M. Lévy-Leblond and F. Balibar, *Quantics: Rudiments of Quantum Physics* (North Holland, Amsterdam, 1990), chapter 1.

[9] This dependence in $\omega^4$ is commonly associated in physics courses with the Rayleigh diffusion and the blue color of the sky.

[10] A different discussion of the gravitational power, involving dimensional arguments, is led in Bernard F. Schutz *Gravity from the Ground Up* (Cambridge: Cambridge University Press, 2003) p 315.

[11] In the general case of a planar trajectory $V^3/R = V a_N = |\mathbf{V} \wedge \mathbf{a}|$. A standard calculation of $\mathbf{V}$ and $\mathbf{a}$ in polar coordinates using the variable $u = 1/r$ leads to $V^6/R^2 = \dot\theta^6 u^{-6}(u+u'')^2$, where $u'' = d^2 u/d\theta^2$ (for Keplerian orbits $u + u'' = \alpha K^{-2}$ is a constant).

[12] See e.g. Lev D Landau and Evgueni M Lifschitz, *Mechanics*, **3**th edition (Pergamon Press, Oxford, 1988), Vol. **1;** Bradley W. Carroll and Dale A. Ostlie *An introduction to modern astrophysics* (Addison Wesley, 2007) 2$^{nd}$ edition.

---

[13] In electromagnetism, for charges involved in harmonic motions ($\omega$ fixed) on elliptical trajectories, $\bar{P}$ is proportional to $a^2 + b^2$, where $b^2 = a^2(1-e^2)$; then $f(e) = 1 - e^2/2$ is a reducing factor.